\documentclass[sn-standardnature]{sn-jnl}

\jyear{2023}%

\usepackage{gensymb, verbatim}

\begin{document}

\title[A TFLN Near-Infrared Platform for Multiplexing Quantum Nodes]{A Thin Film Lithium Niobate Near-Infrared Platform for Multiplexing Quantum Nodes}

\author*[1]{\fnm{Daniel} \sur{Assumpcao}}\email{dassumpcao@g.harvard.edu}\equalcont{These authors contributed equally to this work.}
\author*[1]{\fnm{Dylan} \sur{Renaud}}\email{renaud@g.harvard.edu}\equalcont{These authors contributed equally to this work.}
\author[1]{\fnm{Aida} \sur{Baradari}}
\author[2]{\fnm{Beibei} \sur{Zeng}}
\author[2]{\fnm{Chawina} \sur{De-Eknamkul}}
\author[1]{\fnm{C.J.} \sur{Xin}}
\author[3]{\fnm{Amirhassan} \sur{Shams-Ansari}}
\author[1]{\fnm{David} \sur{Barton}}
\author[2]{\fnm{Bartholomeus} \sur{Machielse}}
\author*[1]{\fnm{Marko} \sur{Loncar}}\email{loncar@g.harvard.edu}

\affil[1]{\orgdiv{John A. Paulson School of Engineering and Applied Sciences}, \orgname{Harvard University}, \orgaddress{\city{Cambridge}, \postcode{02138}, \state{MA}, \country{United States}}}
\affil[2]{\orgdiv{AWS Center for Quantum Networking}, \orgaddress{\city{Boston}, \postcode{02135}, \state{MA}, \country{United States}}}
\affil[3]{\orgname{DRS Daylight Solutions}, \orgaddress{\city{San Diego}, \postcode{16465}, \state{CA}, \country{United States}}}


\abstract{} 
Practical quantum networks will require quantum nodes consisting of many memory qubits. This in turn will increase the complexity of the photonic circuits needed to control each qubit and will require strategies to multiplex memories and overcome the inhomogeneous distribution of their transition frequencies. Integrated photonics operating at visible to near-infrared (VNIR) wavelength range,  compatible with the transition frequencies of leading quantum memory systems, can provide solutions to these needs. In this work, we realize a VNIR thin-film lithium niobate (TFLN) integrated photonics platform with the key components to meet these requirements. These include low-loss couplers ($<$ 1 dB/facet), switches ($>$ 20 dB extinction), and high-bandwidth electro-optic modulators ($>$ 50 GHz). With these devices we demonstrate high-efficiency and CW-compatible frequency shifting ($>$ 50 $\%$ efficiency at 15 GHz), as well as simultaneous laser amplitude and frequency control through a nested modulator structure. Finally, we highlight an architecture for multiplexing quantum memories using the demonstrated TFLN components, and outline how this platform can enable a 2-order of magnitude improvement in entanglement rates over single memory nodes. Our results demonstrate that TFLN can meet the necessary performance and scalability benchmarks to enable large-scale quantum nodes. 



\maketitle

\section{Introduction}\label{sec1}

The distribution of quantum entanglement across continental distances via quantum networks is an enabling technology for a variety of applications, including quantum key distribution, sensing, and distributed quantum computing \cite{Kimble2008, Wehner2018}. An important step towards this goal is generation of remote entanglement between spatially separated quantum memory nodes. Despite great experimental progress \cite{Covey2023, Hensen2015, Hermans2022, Leent2022, Delteil2016, Knaut2023}, the rate of entanglement generation across km scale distances has been limited to the sub-Hz regime. Transitioning these experiments from proof-of-concept demonstrations to practical networking technology requires increasing the number of qubits to 100s-1000s or even beyond. This would enable both increasing entanglement generation rates through multiplexing as well as the utilization of qubits as auxiliary memories for multi-qubit protocols such as entanglement distillation or entanglement swapping \cite{Wehner2018, Duan2001, Bennett1996}. 

To effectively scale the number of qubits per node, however, several key technical innovations are necessary. First, the qubits and associated controls must be manufacturable in a scalable way. While solid-state qubits satisfy this criteria, their optical control currently requires bulky off-the-shelf lasers as well as acousto-optic and electro-optic modulators. Moreover, solid-state qubits, such as color centers, suffer from variations in their transition frequency  which prevents entanglement generation between these qubits \cite{Evans2016}. Therefore, a method to efficiently frequency shift photons emitted by the qubits is required to overcome this inhomegenous broadening. Finally, the ability to connect any of the on-chip memories  to a common optical channel (e.g. optical fiber)  is essential to enable temporal multiplexing that can significantly increase  the entanglement generation rate. This can be accomplished using a large on-chip switch network . All of these functionalities must be implemented with minimal additional insertion loss. This can be challenging since , the majority of quantum emitters have optical transitions in the visible and near-infrared regime (400 nm - 1000 nm) which is above the bandgap of traditional photonic integrated circuit (PIC) platforms such as indium phosphide or silicon \cite{Zeng2023}. On the other hand,  silicon nitride  does offer a suitable transparency window, but available modulation approaches based on the thermo-optic effect or the piezo-electric effect, are slow and cannot achieve 10s to 100s MHz switching speeds required for multiplexing \cite{Mohanty2020, Sacher2019, Dong2022}.

Recently thin-film lithium-niobate (TFLN) on insulator has emerged as a promising PIC platform due to its combination of large electro-optic (EO) coefficient, low optical losses, and large transparency window extending to 400 nm \cite{Zhang2023, Zhu2021}. High-performance visible systems have been fabricated on TFLN including low-loss passive devices and active modulators \cite{Desiatov2019, Renaud2023, Christen2022, Sund2023, Valdez2023}.

In this work we extend these results and  demonstrate the key building blocks needed to scale quantum networking technology. We propose a TFLN photonic enabled architecture to realize a scalable quantum node, and demonstrate the necessary devices to realize this vision. This includes high-efficiency couplers with coupling efficiencies below 1 dB/facet, high-bandwidth phase modulators which we utilize to frequency shift a CW tone by up to 15 GHz with efficiencies beyond $50\%$, and optical switches with high extinction ratio (20 dB) and suitable EO  bandwidth for multiplexing. To both highlight the scalability of the platform and provide a compact solution for qubit optical control hardware, we also demonstrate a multi-component laser control unit PIC, which can be used to provide amplitude and frequency control of an input laser in our proposed architecture. 

\section{Results}\label{sec2}

\begin{figure}[ht]
\centering
{\includegraphics[scale=0.6]{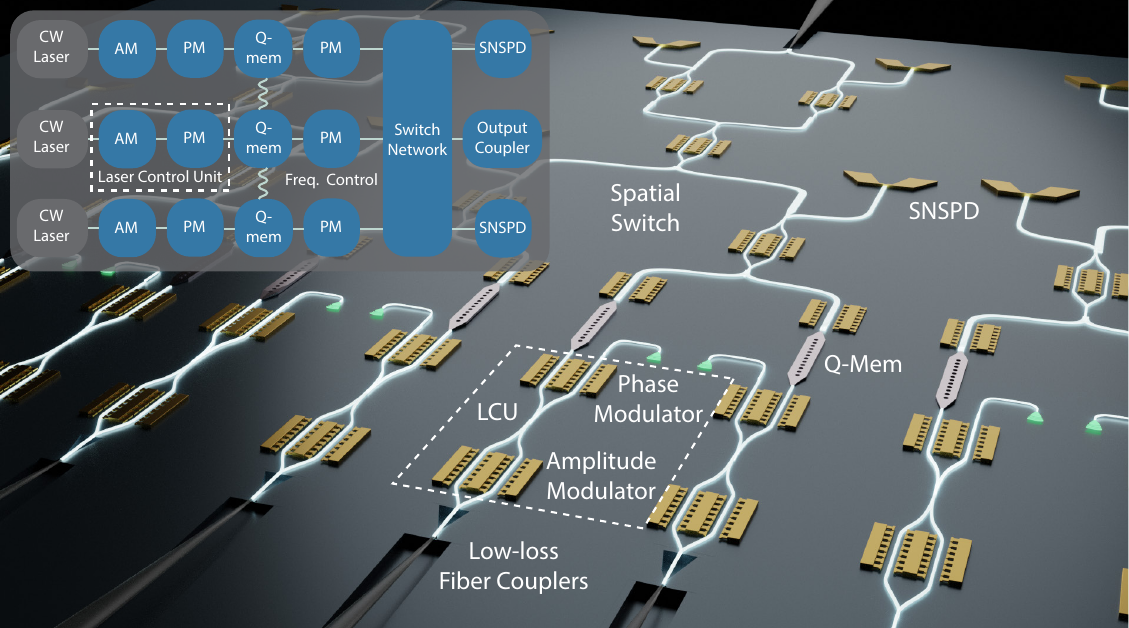}}
\caption{ \textbf{Illustration of proposed TFLN-enabled quantum node}. Light from a laser is fiber-coupled to TFLN PIC and then passes through  a laser control unit (LCU). that consists of an  amplitude modulator (AM) followed by a phase modulator (PM), The LCU provides amplitude and frequency control of the input CW laser, needed to address a quantum memory module (Q-Mem) . We emphasize that while the Q-Mem is shown on chip for illustrative purposes, it can be effectively linked to the main control PIC via optical fiber and low loss couplers demonstrated in this work. Another PM enables frequency control of single photons entangled with the Q-Mem. Finally, a  switch network is used to  multiplex the various memories together. Photons passed through the switch are routed to either on-chip single photon detectors (SNSPDs) or to a low-loss output fiber coupler to route photons to the off-chip fiber network.}
\label{overview}
\end{figure}

Our proposed TFLN architecture to realize a scalable quantum network node is depicted in Fig. \ref{overview}. It consists of a combination of amplitude modulators (AM), frequency shifting via phase modulators (PM), switches, and high efficiency fiber-couplers all interfaced with a large array of quantum memory (QM) devices (see supplementary information for detailed discussion). These devices provide the necessary functionalities to effectively scale the optical control and multiplex the output of the QM modules. We note that although in our schematic the QMs are depicted as integrated within the PIC, this is not a requisite for our architecture. For example, the TFLN control PIC can be connected to QMs on a separate chip using low loss fiber couplers \cite{Zhang2023, Burek2017}, that we discuss next. 

\begin{figure}[ht]
\centering
{\includegraphics[scale=0.6]{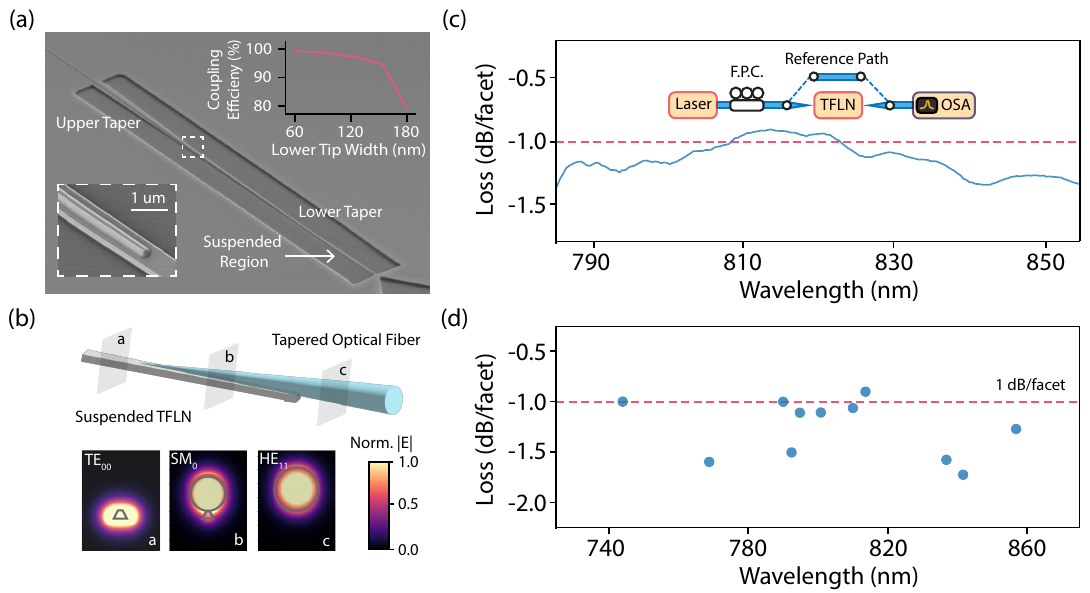}}
\caption{\textbf{High-efficiency tapered optical fiber coupler for VNIR TFLN}. (a) Scanning electron micrograph (SEM) of a free-standing bilayer  adiabatic fiber coupler realized in TFLN. Light from a tapered optical fiber is adiabatically transferred to a ridge waveguide mode of a TFLN "lower taper", and then using "upper taper" is transferred into the rib-waveguide mode.  (lower inset) SEM magnifying the bilayer transition region. (upper inset) Simulated coupling efficiency as a function of lower taper tip width, with a maximum simulated efficiency of $\sim 99\%$. (b) Schematic of the TFLN-fiber interface alongside optical mode simulations at various cross-sections. The simulations illustrate the transition from TFLN waveguide mode to optical fiber mode. (c) Coupling loss measurements, showing a minimum chip-to-fiber coupling loss of 0.9 dB/facet at $\sim 810$ nm. (d) Minimum coupling loss and corresponding minimum loss wavelengths measured across a variety of devices, demonstrating reproducible coupling losses below 1.5 dB/facet.}
\label{coupler}
\end{figure}

For high-fidelity and high-rate  entanglement generation and distribution,  minimizing insertion loss of all optical components, including the on/off-chip coupling losses, is paramount. Inspired by the ultra-low coupling efficiencies achieved using adiabatic tapered fiber-to-chip interfaces in diamond and silicon-nitride platforms, we develop an adiabatic fiber coupler in TFLN (Fig. \ref{coupler}) \cite{Burek2017, Zeng2023, Khan2020, Tiecke2015}. The primary challenge for low-loss adiabatic fiber couplers in TFLN is underlying oxide cladding that can support leaky modes. Therefore, to achieve efficient coupling between the TFLN and optical fiber, the TFLN waveguide must be suspended. We implement our design by first  using a TFLN bilayer taper \cite{He2019} to adiabatically convert the rib waveguide mode to a ridge waveguide. Next, the ridge width is decreased to create an inverted taper ("lower taper") which expands the mode (Fig. \ref{coupler}a) so that it can be coupled to a tapered optical fiber. A clamp is included at the end of the suspended lower taper to prevent collapsing of the device. A tapered optical fiber touches down on the TFLN lower taper, ensuring an efficient, adiabatic mode transfer between the fiber and TFLN waveguides. The geometry was optimized via finite-difference-time-domain (FDTD) simulations, with a peak simulated efficiency of $\sim$ 99$\%$ (Fig. \ref{coupler}a inset). The achievable efficiency in simulations is primarily limited by the lower tip dimension, which in our fabricated devices is $\sim 60$ nm. Fabricated devices (for details, see methods) were characterized and featured coupling losses as low as 0.9 dB/facet (81.2$\%$ efficiency), with optical bandwidths exceeding 100 nm.  The discrepancy between simulated and measured losses is attributed to scattering at the bilayer taper rib-ridge mode conversion interface, shown in the inset of Fig. \ref{coupler}a. Still we emphasize that measured fiber-coupling values are state-of-the-art for TFLN couplers at visible to near-infrared wavelengths, and are comparable to the best demonstrated at telecom wavelengths as well \cite{Lomonte2021, Hu2021}.

\begin{figure}[ht]
\centering
{\includegraphics[scale=0.6]{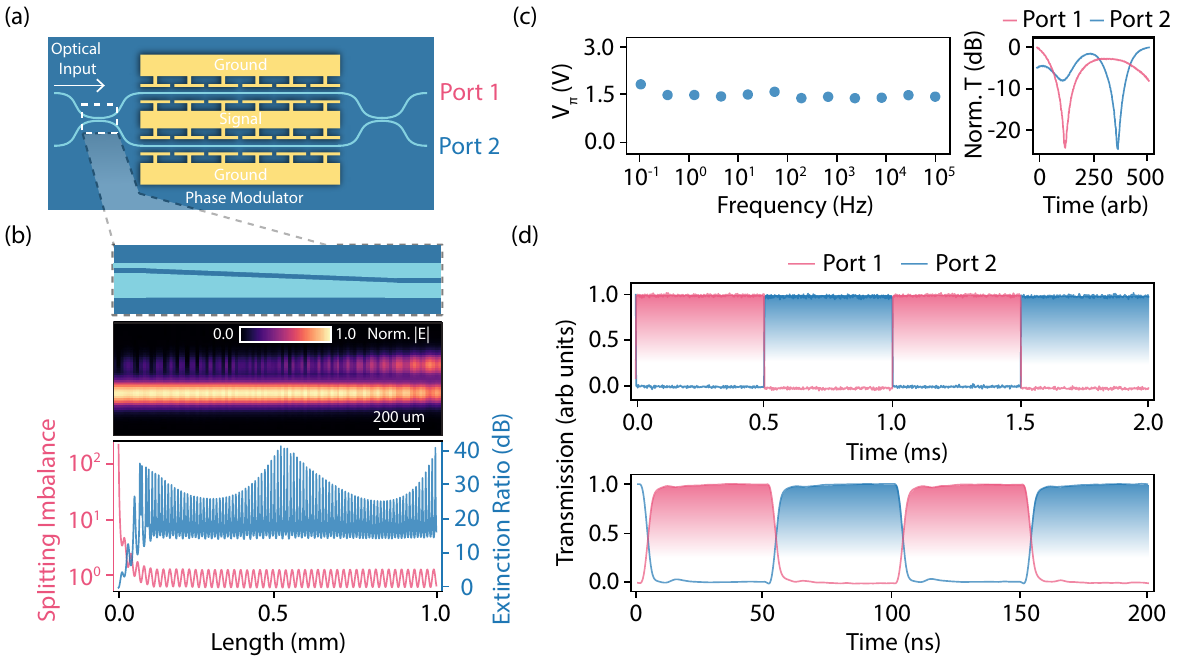}}
\caption{\textbf{TFLN  switches.} (a) Diagram of a TFLN Mach-Zehnder modulator (MZM) based optical switch. (b) (top) Diagram and EME simulation of the adiabatic 2x2 coupler used in the switch. (bottom) Simulated coupler splitting ratio and corresponding modulator extinction ratio as a function of length. For lengths longer than the adiabaticity criterion ($> 100$ \micro m), we observe a reduced dependence of split ratio and extinction on device length, as expected. (c) Experimentally measured 5 mm long MZI switch features  (left) a flat $V_\pi$ of 1.5V and (right) extinction ratio $ > 20$ dB for both ports. (d) Switching demonstration of the device, showing stable switching at both kHz and MHz time scales.}
\label{switch}
\end{figure}

Spatial switches are required to route photons for both multiplexing and on-chip entanglement generation. We implement an EO switch using a Mach-Zehnder modulator (MZM) based design with 2x2 couplers at the input and output of the device (Fig. \ref{switch}a). Adiabatic couplers are used for the 2x2 splitter to ensure fabrication robustness (Fig. \ref{switch}b) (see supplementary information for additional design details) \cite{Cao2010, Yun2015}. We use numerical simulations to evaluate the splitting imbalance  of the coupler,  defined as the ratio of powers at the two output ports when coupler is excited using  a single input port. We also evaluate the maximum extinction ratio achievable by a Mach Zehnder modulator utilizing two of these couplers due to the simulated splitting imbalance (Fig. \ref{switch}b). Although some the splitting ratio and corresponding modulator extinction ratio still depends on the length of the coupler, likely due to partial coupling of the two waveguides on the imbalanced width side of the splitter, splitting imbalances are still close to 1 and extinction ratio of 15dB can still be achieved even in the worst case. This indicates tolerance to fabrication imperfections. Fabricated devices feature an on-chip loss of 0.4 dB per switch, which is limited by on-chip propagation losses (see supplementary information for supporting measurements). To control the switch, we employ capacitive lumped element electrodes (as opposed to traveling-wave designs discussed below) to limit power consumption. A half-wave voltage ($V_\pi$) of $\sim 1.5$V is measured for a 5 mm modulator, with extinction ratios greater than 20dB on both ports (Fig. \ref{switch}c). We note that the observed $V_\pi$ is flat as a function of frequency down to 0.1 Hz, which is important both for low frequency switching operation and achieving a stable modulator bias \cite{Holzgrafe2024}. We leverage this flat response to demonstrate switching between the two ports at kHz and MHz timescales, both of which are relevant for memory and photon operations, respectively, for typical solid-state emitters (Fig. \ref{switch}d) \cite{Nguyen2019, Hensen2015, Lukin2020, Bradac2019}. Importantly, these capacitive electrodes do not dissipate significant power in the hold state , and only  $\sim$100$\micro W$ for 10 MHz switching, (see supplementary information for details). This suggests that large switch networks can be created with minimal electrical overhead, making them suitable for multiplexing quantum memories even at cryogenic temperatures.

\begin{figure}[ht]
\centering
{\includegraphics[scale=0.6]{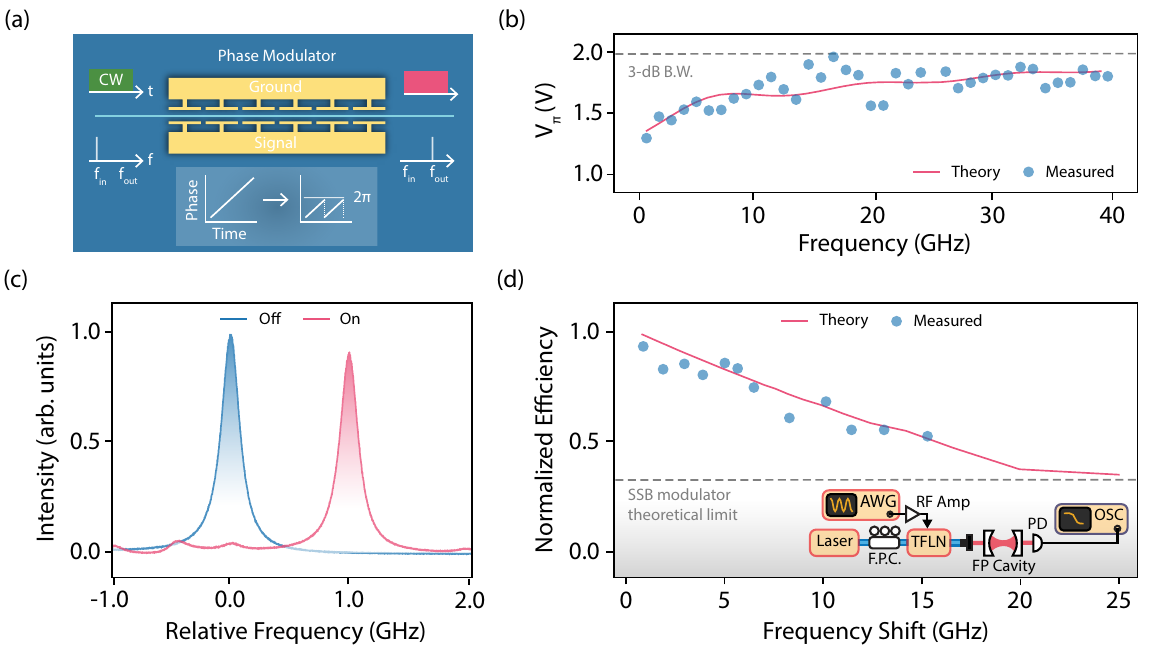}}
\caption{\textbf{High-Bandwidth Phase Modulator for Frequency Shifting}. (a) Schematic of phase modulator utilizing segmented electrodes to achieve high-bandwidth EO performance. An input optical signal is frequency shifted using EO serrodyning, where a linear increase in optical phase is approximated via a periodic sawtooth wave whose amplitude corresponds with a phase of 2$\pi$. (b) Half-wave voltage ($V_\pi$) as a function of frequency for a 1 cm phase modulator, showing a 3-dB EO bandwidth exceeding 40 GHz ($~100$ GHz simulated). (c) Frequency spectrum of light transmitted through the phase modulator with the serrodyne electrical signal turned off (blue) or on (red), demonstrating a high-efficiency ($> 90 \%$) 1 GHz frequency shift. (d) Serrodyne shift efficiency as a function of frequency shift, showing efficiencies greater than 50$\%$ for frequency shifts up to 15 GHz.}
\label{serrodyne}
\end{figure}

To overcome the inhomogeneous distribution of QM optical transitions, our platform incorporates  optical frequency-shifting capabilities. We choose to utilize a serrodyne-based frequency approach whereby a linear advance in phase is emulated using a sawtooth wave with an amplitude of $2\pi$ applied to the phase modulator. This approach in principle can enable frequency shifts with unity efficiency and functions irrespective of the temporal shape of the input light (Fig. \ref{serrodyne}a) \cite{Holland2021, Johnson2010, Houtz2009, Poberezhskiy2005, Sinclair2014, Saglamyurek2014} (see supplementary information for details). The latter is especially important, as quantum memories can possess narrow transitions with correspondingly long photon lifetimes (ns or longer). The challenge with serrodyning is that due to the high-bandwidth nature of an electrical sawtooth wave, a large analog bandwidth is required from both the modulator and the control electronics. 

To realize this, we utilize a segmented-electrode PM \cite{Kharel2021}. This design increases the microwave index of the electrode, allowing us to achieve microwave-optical velocity matching with a smaller impedance penalty than achieved previously (see supplementary information for supporting data and simulations). In addition, this electrode design exhibits decreased Ohmic losses due to the breaking of the current path on the inner portion of the electrode \cite{Kharel2021, Ding2014, Shin2011}. With this design, we realize high-bandwidth phase modulators with estimated 3-dB bandwidths of 55 GHz ($>$ 100 GHz) with respect to  2GHz ( 5 GHz) response,  which is a considerable enhancement over previously achieved results (35 GHz with respect to 3 GHz response) \cite{Renaud2023} (Fig. \ref{serrodyne}b). We estimate an on-chip optical loss of 0.7 dB for a 1-cm device (see supplementary information for supporting measurements). 

Utilizing this modulator we perform a proof-of-principle serrodyne frequency shifting demonstration. We apply a sawtooth electrical waveform onto the phase modulator using a fast arbitrary waveform generator (AWG) and radio-frequency (RF) amplifier. The transmitted light is measured in the frequency domain by a scanning Fabry-Perot (FP) cavity. We observe frequency shifts of 1 GHz with high ($ > 90 \%$) efficiencies (Fig. \ref{serrodyne}c). To understand the limits of this serrodyning technique, we sweep the shift frequency and measure the efficiency (Fig. \ref{serrodyne}d). We find that there is an efficiency roll-off at higher frequencies attributed to the limited  analog electrical bandwidth of 35 GHz  amplifier used in our experiments. A model considering the finite bandwidth of the amplifier accurately reproduces our results, as shown in Fig. \ref{serrodyne}d (see supplementary information for model information). Despite this, we are still able to achieve shifts up to 15 GHz with shift efficiencies beyond 50$\%$. Notably, this is sufficient to overcome the inhomogeneous distribution of various diamond QMs \cite{Evans2016, Machielse2019}, and a significant improvement over previous demonstrations which were limited to $\pm$ 5 GHz \cite{Sinclair2014, Saglamyurek2014}.  Going forward, we note that using state-of-the-art high-bandwidth electronics, even larger shifts can be achieved using the same modulator (see supplementary information for details) \cite{Chen2017}. This  demonstrates the utility of our PMs coupled with  serrodyning technique to achieve ultra-large frequency shifts, with small added loss, for overcoming the inhomogeneous distribution of solid state QMs.

\begin{figure}[ht]
\centering
{\includegraphics[scale=0.6]{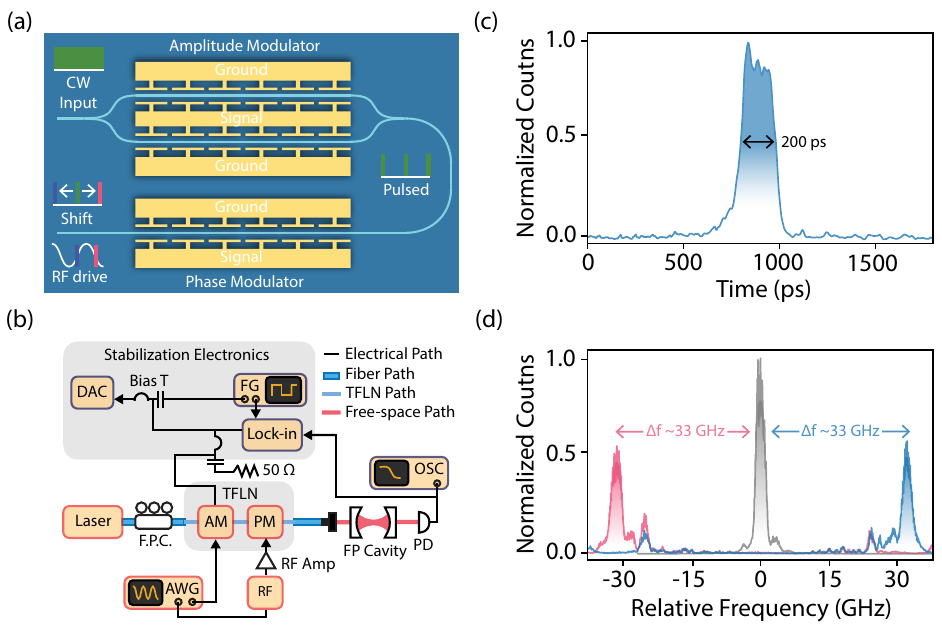}}
\caption{\textbf{Laser control unit (LCU).} (a) Schematic of LCU, that combines  an AM  to form optical pulses, and PM for frequency shifting via optical shearing effect  of laser light. (b) Diagram of the experimental setup used to characterize LCU. (c) Temporal profile of the generated pulse with pulse-width down to 200 ps. (d) Frequency spectrum of generated and frequency shifted optical pulses. A shift of $\pm$ 33 GHz is demonstrated. }
\label{LCU}
\end{figure}

To control QMs, CW and pulsed laser signals are needed. Inhomogeneous broadening of QMs, however, requires laser light of different wavelengths and pulse durations. One approach would be to utilize integrated photonic lasers with emission wavelengths tuned to particular QMs of interest \cite{Tran2019, amirlaser}.  This approach does not scale well, as it would require an individual laser per quantum memory. QM optical control typically requires low laser linewidth for high-fidelity operations (kHz) and low optical power. As laser optical power and laser linewidth are inversely correlated \cite{Henry1982, Coldren2012}, it is most efficient to have one integrated, higher power and low linewidth laser driving multiple QMs. This can be achieved using a laser control unit (LCU, Fig. \ref{overview}) consisting of an AM followed by a PM (Fig. \ref{LCU}a) for each QM. This combination not only enables the amplitude and frequency shaping of input CW light to generate optical control pulses to interface with QMs, but also shows the ability of our platform to support complex PICs containing multiple nested devices.

To illustrate the capabilities of the LCU, we implement a pulse shaping and shearing protocol to create a train of short pulses, and subsequently shift their central frequency. The AM is used to carve short pulses, which are then sent to a PM and temporally aligned with the linear rising/falling region of an applied sinusoidal tone (Fig. \ref{LCU}a). This introduces a frequency shift, and is commonly referred to as EO shearing \cite{Johnson1988, Wright2017, Zhu2022}. We have previously demonstrated state-of-the-art EO shearing with TFLN modulators, measuring maximum shifts of over 7 times the pulse linewidth \cite{Renaud2023}. However, in that work the pulse generation was done with a separate commercial AM. As the LCU contains both an AM and PM, we are able to perform amplitude and frequency shaping operations within the same PIC. 

We utilize a electronic setup similar to our previous demonstration, with both the sinusoidal tone and the electrical pulses originating from a common AWG. The sinusoidal tone and RF pulses are applied to the PM and AM, respectively (Fig. \ref{LCU}b). We bias the amplitude modulator at the null point for pulse generation using a quasi-DC electro-optic bias whereby a feedback loop is applied to the low-frequency electrical bias of the modulator loop to overcome the DC bias drift prevalent in TFLN modulators \cite{Sosunov2021, Holzgrafe2024}. By avoiding biasing using heaters, we ensure our devices can be used in cryogenic settings (see supplementary information for additional details). 

Short (200 ps) optical pulses are generated by the AM directly driven by the AWG (Fig. \ref{LCU}c), which is faster than typically required for quantum optical control pulses. We then apply a 250 MHz sinusoidal tone to the PM and measure the resultant frequency spectrum via a FP cavity (Fig. \ref{LCU}d). We find that we can deterministically shift our optical pulses by $\pm 33$ GHz. This corresponds to a drive amplitude of 40$V_\pi$ and a shift of over 16 times the linewidth (see supplementary information for details). This large shift is sufficient to cover the inhomogeneous distribution of a variety of different emitters, providing a method for creating optical pulses to address different emitters, all while being seeded by a single laser.  We observe some distortion is induced via shearing, as seen in the spectrum of the sheared transmission spectra. We hypothesize that this distortion is due to the inherent nonlinearity of the sinusoidal drive signal used in our experiment (see supplementary information for details on the model). We believe that LCU will be crucial to limit the complexity of optical control while scaling up the number of qubits per node. Through direct integration of the laser within the PIC, its scalability can be enhanced even further \cite{amirlaser}. Beyond this, these findings demonstrate the ability to create complex, multi-element PICs with our platform. 

\section{Discussion}\label{sec3}

\begin{figure}[ht]
\centering
{\includegraphics[scale=0.6]{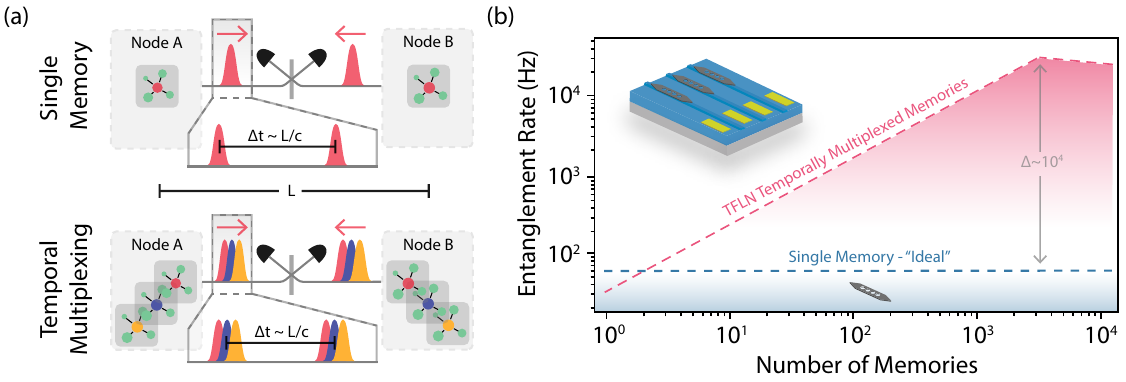}}
\caption{\textbf{TFLN Enabled Temporal Multiplexing Advantage} (a) Diagram of remote entanglement generation scheme between two quantum nodes separated by a distance L (top)  with a single memory or (bottom) using multiple memories interleaved via temporal multiplexing. (b) Computed entanglement rate between two quantum memories separated by L = 20km utilizing temporal multiplexing as a function of number of memories used. Losses of our platform are incorporated in this model. Due to the low-loss and scalability of our TFLN devices, a gain of over 100x in entanglement rate can be achieved via temporal multiplexing as compared to the single memory case. }
\label{multiplex}
\end{figure}

In this work we have demonstrated the necessary ingredients to create PICs for multiplexing quantum network nodes, including state-of-the-art modulators, switches, and high efficiency couplers. Using this platform we have shown proof of principle demonstrations of its utility, including frequency shifting a CW tone beyond 15 GHz with efficiencies above 50$\%$ and creating multi-element PICs to provide amplitude and frequency control of optical pulses for simplifying optical control complexity at scale.

To contextualize the potential of this platform, we compute the benefits of temporal multiplexing using the TFLN-based architecture in Fig. \ref{overview} and our demonstrated device performance. We consider the model of two quantum nodes separated by a distance $L$, where each node probabilistically generates a flying photon entangled with the local spin memory and sent to a central station where a detection of one of the photons heralds entanglement (see supplementary information for details) (Fig. \ref{multiplex}). Although this architecture is largely quantum memory agnostic, for discussion purposes we  focus on solid-state defects in diamond, and specifically the silicon vacancy center in diamond, due to their excellent optical and spin properties and ability to be fabricated at scale \cite{Evans2016, Tchernij2017, Lukin2020, Doherty2013, Bhaskar2020}. We consider two cases. The first is where only a single memory exists at each quantum node, and each node must wait the classical signaling time ($L/c$) between each entanglement attempt, limiting the repetition rate. The second case is where each quantum node has multiple memories which can be temporally interleaved via our TFLN PIC. This allows a significant increase in the repetition rate limited by the number of memories at each node and ultimately by the photon length,at the cost of additional losses from the TFLN PIC including coupling losses, frequency shifting losses, and device losses. We compute the achievable entanglement rates across a 20km channel for these two cases utilizing a silicon-vacancy center in diamond based quantum memory. In the temporal multiplexing case we incorporate the estimated losses of the PIC including both individual device loss and off-chip coupling losses (see supplementary information for full model details). Despite the additional overhead of the TFLN PIC, we estimate the advantage of temporal multiplexing will lead to an enhanced entanglement rate with even a modest number of memories ($<$ 10) with an over 100x increase in rate achievable through saturation of the channel temporally (requiring $\sim 2000$ memories) (Fig \ref{multiplex}b). Through additional wavelength division multiplexing (WDM) using current telecom WDM technology, we estimate MHz entanglement rates can be achieved across the 20 km distance. This therefore demonstrates our platform achieves the necessary performance metrics and scalability to enable multiplexing and maximize quantum network performance.

Using this platform, future work will focus on increasing the complexity of visible TFLN PICs through increasing the number of elements and integrating with quantum memory systems to demonstrate the advantage of scalability. Moreover, due to the state-of-the-art performance of these devices, this platform can be applied to other quantum technology applications, such as control of atomic platforms, and even classical applications such as AR/VR, sensing, or communications operating in the visible region.

\section{Methods}\label{sec4}
\subsection{Device fabrication}
All TFLN devices were realized on  thin film wafers from NanoLN (X-cut, 300 nm LN/7 $\mu$m SiO$_2$/500 $\mu$m Si). Integrated photonic devices are fabricated by first patterning waveguide layers using electron-beam lithography (HSQ, Elionix ELS-F125), and etching $\sim$180 nm using reactive ion etching (Ar$^+$). This is then followed by re-deposition cleaning (RCA SC-1) and a high-temperature anneal (520 $\degree$C).  For fabrication details please see Ref \cite{Renaud2023}.

\subsection{Modulator Measurements}
For low frequency measurements, a variable frequency function generator produces a triangular waveform that is directed to the MZM electrodes using a 50 $\Omega$ ground-signal-ground probe (40A-GSG-100-F). A CW laser ($\lambda = 730$) is  coupled to the modulator and out-coupled to an avalanche photodetector (APD410A). The detector output is monitored on an electronic oscilloscope, from which the $V_{\pi}$ is extracted. 

For high frequency $V_{\pi}$ measurements, a sinusoidal RF tone is applied at varying frequencies and the resulting optical spectrum is measured. The optical frequency spectrum of an MZM given an input CW optical carrier frequency of $\omega_0$, an applied RF tone at frequency $\omega_m$ and amplitude $V_0$, and internal phase between the arms of $\phi$ is given by:

\begin{equation}
    I(\omega_0 + k\omega_m) \propto \frac{1}{2}J_k^2(\pi V_0/V_\pi)[1 + (-1)^k\text{cos}(\phi)]
\label{eq:refname1}
\end{equation}
where $k$ is an integer of the harmonic of the drive frequency \cite{Shi2003}.

The frequency spectrum is measured using a home-built Fabry-Perot cavity (linewidth = 200 MHz) and fit to \ref{eq:refname1}.

For loss measurements, propagation loss was assumed to be the dominant loss, which was extracted via cutback measurements to be 0.7 dB/cm \cite{Renaud2023}. 

\subsection{Serrodyne Demonstration}
Sawtooth electrical waveforms were generated using a fast arbitrary waveform generator (92 GSa, Keysight) and amplified via a high-bandwidth high-power amplifier. Pre-compensation is applied by the AWG to compensate for frequency-dependent loss induced distortions from both the cables, and the amplifier itself. The frequency spectrum of light transmitted through the modulator was measured via a scanning Fabry-Perot cavity. To compute the shift efficiency, the frequency spectrum of light transmitted through the modulator was measured with and without the applied sawtooth in quick succession to limit any setup drifts. The ratio of power in the shifted sideband to the initial unperturbed power was taken as the efficiency. 

\subsection{Shearing Demonstration}

To bias the amplitude modulator at the null point, a dither-based feedback was used. A small sinusoidal frequency at a low frequency (1 kHz) was combined with a DC offset via a bias tee and then applied to the amplitude modulator via a second bias tee. The transmitted light was then detected on a photodetector and the voltage output routed to a lock in amplifier to demodulate the dither signal and provide an error value. A software based PID loop was then used to feedback the DC bias on this error value and ensure the amplitude modulator was locked to the null point. 

With the amplitude modulator biased at the null point, short electrical pulses were applied to the amplitude modulator to generate the optical pulses while a sinusoidal tone was applied to the phase modulator. Both were generated by the same AWG (Tektronix 70000). For the shearing demonstration, an optical pulse length of 400 ps was used. Although shorter pulses could be generated, the total frequency shift with these pulses would be larger than the free-spectral range of the FP cavity (70 GHz) leading to aliasing artifacts. 

We also note that the phase modulator was operated in lumped element configuration with no termination. This was possible due to the electrical narrowband nature of shearing. 
\subsection{Adiabatic Coupler Design}
We simulate this coupler using an eigenmode expansion (EME) solver (Lumerical) to calculate the splitting ratio and the correspond extinction ratio of the modulator as a function of length. We find that the coupler reaches the adiabatic condition at a length of 100 \micro m. While there is still a variation of splitting ratio and extinction for longer lengths, the worst case extinction ratio is still 15 dB, thus ensuring fabrication robustness. The simulated insertion loss is negligible ($\sim 0.01$ dB).

\backmatter

\bmhead{Supplementary information} See supplementary document for supporting content. 

\section*{Declarations} 
\bmhead{Acknowledgements} This work was supported in part by AFOSR FA9550-20-1-0105 (M.L., D. R.), AFOSR FA9550-19-1-0376 (M.L., A. S. A), ARO MURI W911NF1810432 (D. R. D. B., M.L.), NSF EEC-1941583 (M.L. D. A. K. P., C. X.), NSF OMA-2137723 (M.L., C. X.), OMA-2138068 (M.L., M. Y.), AWS Center for Quantum Networking’s research alliance with the Harvard Quantum Initiative (M.L. D. A. D. R., C. X.), Ford Foundation Fellowship, (D.R.), and NSF GRFP No. DGE1745303 (D.R., D.A.). Device fabrication was performed at the Center for Nanoscale Systems (CNS), a member of the National Nanotechnology Coordinated Infrastructure Network (NNCI), which is supported by the National Science Foundation under NSF Grant No. 1541959.
\bmhead{Competing interests} M.L. is involved in developing lithium niobate technologies at HyperLight Corporation. The remaining authors declare no competing interests
\bmhead{Availability of data and materials} The data presented in this study is available from the corresponding authors upon request.
\bmhead{Authors’ contributions} These authors contributed equally: Daniel Assumpcao and Dylan Renaud.

\bigskip

\newpage

\bibliography{sn-bibliography-upd}

\end{document}


\title[A TFLN Near-Infrared Platform for Multiplexed Quantum Nodes]{A Thin Film Lithium Niobate Near-Infrared Platform for Multiplexed Quantum Nodes}

\author*[1]{\fnm{Daniel} \sur{Assumpcao}}\email{dassumpcao@g.harvard.edu}\equalcont{These authors contributed equally to this work.}
\author*[1]{\fnm{Dylan} \sur{Renaud}}\email{renaud@g.harvard.edu}\equalcont{These authors contributed equally to this work.}
\author[1]{\fnm{Aida} \sur{Baradari}}
\author[2]{\fnm{Beibei} \sur{Zeng}}
\author[2]{\fnm{Chawina} \sur{De-Eknamkul}}
\author[1]{\fnm{C.J.} \sur{Xin}}
\author[3]{\fnm{Amirhassan} \sur{Shams-Ansari}}
\author[1]{\fnm{David} \sur{Barton}}
\author[2]{\fnm{Bartholomeus} \sur{Machielse}}
\author*[1]{\fnm{Marko} \sur{Loncar}}\email{loncar@g.harvard.edu}

\affil[1]{\orgdiv{John A. Paulson School of Engineering and Applied Sciences}, \orgname{Harvard University}, \orgaddress{\city{Cambridge}, \postcode{02138}, \state{MA}, \country{United States}}}
\affil[2]{\orgdiv{AWS Center for Quantum Networking}, \orgaddress{\city{Boston}, \postcode{02135}, \state{MA}, \country{United States}}}
\affil[3]{\orgname{\tex{DRS Daylight Solutions}}, \orgaddress{\city{\tex{San Diego}}, \postcode{\tex{16465}}, \state{\tex{CA}}, \country{\tex{United States}}}}

\keywords{Visible Wavelength, Lithium Niobate, Integrated Photonics}



\maketitle

\section{Electro-Optic Shifting}
Electro-optic frequency shifting of photons is a possible solution for the inhomogenous distribution of solid-state emitters. 

Phase modulators can in principle shift the frequency of CW optical tones with unity efficiency through the application of a linear phase advance or retardation to the transmitted field. More formally, if am input CW tone $Ae^{j\omega t}$  with frequency $\omega$ is transmitted through a phase modulator with an applied phase $\phi (t)$, the transmitted field is given by $E_{T, ideal}(t) = Ae^{j(\omega t + \phi (t))}$. When $\phi (t) = kt$ we observe that $E_{T, ideal}(t) = Ae^{j(\omega + k)t} = Ae^{j\omega '}$: a CW tone with a shifted frequency $\omega ' = \omega + k$

In practice, a limitless linear increase in phase would require a correspondingly unbounded increase in applied voltage which is nonphysical. Thus various techniques are used to emulate a linear phase advance. In this work we explore two techniques: shearing and serrodyning. In shearing, a sinusoidal tone is applied to the phase modulator and an optical pulse is co-localized to the linear region of the phase to emulate a linear rise. This technique requires the input light be pulsed, thus causing a fundamental trade-off between the pulse linewidth and frequency shift. With serrodyning, an electrical sawtooth wave is used with an amplitude of 2$\pi$. As a 2$\pi$ phase is equivalent to 0 phase, an ideal sawtooth wave is equivalent to a monotone linear increase. This method works regardless of the initial bandwidth of the input tone, but requires higher bandwidth electro-optic control. 

In this section, we will model both techniques to understand their limitations.

\subsection{Modeling and Limits of Spectral Shearing}

With the shearing technique, a sinusoidal tone with frequency $f_{RF}$ and modulation depth $\phi_0$ is applied to the modulator $$ \phi (t) = \phi_0 sin(2\pi f_{RF}t)$$ To emulate a tone, the input optical signal is assumed to be pulsed with pulses co-localized to the rising edge: $E_{T, ideal}(t) = A(t)e^{j(\omega t + \phi (t))}$. Thus we can perform a taylor expansion: $$ \phi (t) \approx 2\pi f_{RF}\phi_0 t - \phi_0 1/6 (2\pi f_{RF})^3 t^3 + ...$$ where we observe that to first order, an optical frequency shift occurs: $\Delta \omega_{op} = \phi_0 / t = 2\pi f_{RF} \phi_0$. It is clear that in order to maintain the approximate linearity of $\phi (t)$, it is vital that the optical pulse is limited temporally to mitigate the influence of the $t^3$ and higher order terms. Higher order terms will lead to a shift to other frequency bins, which can be considered to be either a  shift distortion or inefficiency.

 There are two figures of merit for shearing: the achievable frequency shift and the efficiency of that shift. The first is given by $\Delta \omega_{op}$ divided by the spectral linewidth of the input pulse $\Delta \nu_{pulse}$, which we define to be the full-width half max in the spectral domain for the purposes of this analysis. The second is the efficiency or distortion of the shift. As shearing is not necessarily periodic and thus the transmission optical spectrum is not intrinsically band-limited, the exact useful definition of efficiency is application dependent. For purposes of this analysis, we define efficiency as the overlap between the transmitted frequency shifted pulse and an ideal frequency shifted pulse $\epsilon = \int{E_T (t) E_{T, ideal}^{*}(t)dt}$.  

Two control variables can be utilized to maximize the performance of shearing, $f_{RF}$ and $\phi_0$. $f_{RF}$ is principal set as a function of the temporal length, $T$, of the pulse that one wishes to frequency shift. If the RF period $1/f_{RF}$ is small compared to $T$, the pulse will experience a larger nonlinear segment of the sinusoidal tone ($t^3$ terms in the Taylor expansion) and thus experience a distortion. This is also why the ratio $\Delta \omega_{op} / \Delta \nu_{pulse}$ is taken as the figure of merit as opposed to the absolute shift $\Delta \omega_{op}$, since in principle the optical pulse can be shrunk to increase the absolute magnitude of the shift, but this will increase the $ \nu_{pulse}$ and thus the ratio provides a fairer metric. 

\begin{figure}[ht]
\centering
{\includegraphics[scale=0.2]{SI Figs/shear_vs_period.pdf}}
\caption{\textbf{Shearing vs RF frequency. } Optical frequency spectra of the transmitted light  for various RF shearing frequencies. We see that when $f_{RF}$ is larger a larger shift occurs, but the shape of the line in the optical frequency domain becomes distorted. (b) Optical frequency shift $\Delta \omega$ versus applied RF frequency, showing an increased achieved shift with a higher RF frequency. (c) Shift efficiency versus applied RF frequency, showing a decreased efficiency at higher RF frequencies due to increased nonlinear distortions. }
\label{ShearingFrequency}
\end{figure}

To numerically evaluate this effect, we compute the transmitted frequency spectra of $E_T (t)$, for a fixed $\phi_0 = 20\pi$ and a square wave optical input with temporal duration $T$ and for a variable RF frequency $f_{RF}$ (Fig \ref{SheaeringAmplitude}). While $f_{RF} = 0.2/T$ achieves double the frequency shift of the case of $f_{RF} = 0.1/T$, it is significantly more distorted. To quantify this trade off $\Delta \omega_{op} / \Delta \nu_{pulse}$ and the efficiency are computed as a function of $f_{RF} T$, thus showing a fundamental tradeoff between the two. As a compromise, for the experiment in this work we pick $f_{RF} = 0.1/T$ as this ensures we have maximized $f_{RF}$ without any significant penalty to the efficiency. 

\begin{figure}[ht]
\centering
{\includegraphics[scale=0.25]{SI Figs/shear_vs_amplitude.pdf}}
\caption{\textbf{Shearing vs drive amplitude}. Optical frequency spectra of the transmitted light  for various phase amplitudes $\phi_0$. We see that increased $\phi_0$ leads to an increased optical shift but also increased distortions. (b) Optical frequency shift $\Delta omega$ versus $\phi_0$ , showing an increased achieved shift with larger amplitudes (c) Shift efficiency versus $\phi_0$, showing a decreased efficiency at higher drive amplitudes due to increased nonlinear distortions. }
\label{SheaeringAmplitude}
\end{figure}

In terms of the effect of the phase modulation index $\phi_0$, one might intuitively think that the goal should be to maximize $\phi_0$ based on the practical limit of applicable RF voltage and minimizing $V_\pi$ of the modulator. Instead we observe that increasing  $\phi_0$ itself also non-negligibly leads to distortion in the signal (Fig \ref{SheaeringAmplitude}). Indeed, numerically we find for the shearing parameters used in this work, if we wanted to further increase $\phi_0$ to achieve $\Delta \omega_{op} / \Delta \nu_{pulse} > 100$ would induce a significant penalty to the efficiency. To understand this, we see that in the taylor expansion of $\phi (t)$, the highest order error term is $\phi_0 1/6 f_{RF}^3 t^3 \propto \phi_0$,. This thus provides another challenge in achieving larger shifts with shearing than just increasing the voltage, and makes it challenging to achieve  $\Delta \omega_{op} / \Delta \nu_{pulse} > 100$ while simultaneously achieving high efficiencies. 

\begin{figure}[ht]
\centering
{\includegraphics[scale=0.3]{SI Figs/shear_efficiency_vs_shift.pdf}}
\caption{\textbf{Efficiency vs Optical Shift.} Efficiency versus normalized optical shift for two different frequencies $f_{RF}$. Shift efficiencies above 90$\%$ for optical shifts $\Delta \omega_{op} / \Delta \nu_{pulse} > 50$ are theoretically achievable using a low $f_{RF}$. }
\label{ShearingEffVsShift}
\end{figure}

There are several possible ways to get around this. First we note that while the error term $\phi_0 1/6 f_{RF}^3 t^3$ has a linear dependence on $\phi_0$ it has a cubic dependence on $f_{RF}$. Thus this term can theoretically be decreased through decreasing $f_{RF}$ and correspondingly using a larger $\phi_0$ (Fig \ref{ShearingEffVsShift}), with the clear disadvantage that an even larger $\phi_0$ is required. To contextualize this, we estimate that to achieve a shift $\Delta \omega_{op} / \Delta \nu_{pulse} > 100$ with efficiency $> 90 \%$ would require $f_{RF} = 0.04/ T$ and $\phi_0 = 160\pi$, corresponding to a voltage amplitude of $\approx 240 V$ for the phase modulators used in this work.  As an alternative strategy, one could use a more complex waveform consisting of the summation of multiple sinusoidal harmonics whose taylor expansion has higher order cancellation: e.g.  $ \phi_2 (t) = \phi_0 [sin(2\pi f_{RF}t) + 1/8 sin(2\pi (2f_{RF})t)] \approx \phi_0 [3/4 (2\pi f_{RF}) t - 1/40 (2\pi f_{RF})^5 t^5 + ... ]$. These more complex waveforms can also be understood as a higher harmonic approximation to a sawtooth wave and thus as a sort of hybrid between serrodyne and shearing, with the disadvantage the additional complexity required as well as a larger $\phi_0$ required to achieve a given shift. 

\subsubsection{Suitability of Shearing for Multiplexing Quantum Nodes}
In the experimental results shown, we have shown the ability to perform high efficiency shearing for $\Delta \omega_{op} / \Delta \nu_{pulse} > 10$. For compensating for the inhomogenous distribution for photons that have been emitted and entangled with the quantum memories, this is insufficient due to the large mismatch between photon lifetimes and inhomogenous distributions. For example, the silicon vacancy center in diamond has a intrinsic linewidth of $\sim$100 MHz but an inhomogenous distribution on the order of $10$ GHz and thus requires $\Delta \omega_{op} / \Delta \nu_{pulse} > 100$.  Given the analysis in the above section, it is clear that increasing this to shifts beyond $\Delta \omega_{op} / \Delta \nu_{pulse} > 100$ will require a greater than 10x increase in the achievable $phi_0$ to compensate for the additional distortions introduced by increasing the drive voltage. Thus it does not seem practical to use shearing for frequency shifting the transmitted entangled single photon signals without significant advances in modulators (potentially using resonant modulator designs or other advances) or on the emitter side to decrease the inhomogenous distribution. Despite this, frequency shearing on the input control side to use a common laser to address a large number of color centers with varying frequencies seems feasible. This is especially true for emission based schemes where excitation optical pulses much shorter than the emitted photon lifetime are used to excite quantum memories and thus have significantly relaxed $\Delta \omega_{op} / \Delta \nu_{pulse}$ requirements and less efficiency and purity requirements as well.

\subsection{Modeling and Limits of Serrodyning}
In serrodyning, a sawtooth wave with an amplitude of $2\pi$ and frequency $f_{ST}$ is applied to the modulator. This emulates a linear wave as a phase of $2\pi$ is equivalent to a phase of $0$ so assuming the drop off region of the sawtooth is instantaneous (as in an ideal sawtooth) the sawtooth is equivalent to a linear increase in phase, and nominally frequency shifts the input light by a frequency $f_{ST}$. Since no temporal pulsing is required, there is no intrinsic trade off between achievable frequency shift and initial pulse linewidth that is observed with shearing -  even a CW tone can be efficiently serrodyned.

The challenge with serrodyning is the generation and signal integrity of the electrical sawtooth wave. The sawtooth wave is a intrinsically high-bandwidth signal, which can be observed either in the time domain due to the sharp drop in voltage at the teeth of the sawtooth, or in the frequency domain where the sawtooth wave can be decomposed into a Fourier series -  $$\phi (t) = \phi_0 [1/2 - 1/\pi \sum_{n=1}^{\inf} \frac{1}{n} sin(n\pi t\cdot f_{ST})]$$ where $\phi_0$ is the modulation depth, and $f_{ST}$ is the frequency of the sawtooth wave. We observe a very modest roll-off $\propto 1/n$ in the Fourier component amplitude thus requiring a large bandwidth to faithfully reproduce the sawtooth wave. Distortions in the sawtooth wave when applied to the modulator causes frequency content to partially shit to other undesired frequency bins, thus decreasing the efficiency. 

For generation of the sawtooth wave, we choose to use a digital approach using a high-speed arbitrary waveform generator (AWG). This is advantageous compared to the previously utilized analog approaches since the AWG allows arbitrary modification of the signal in order to compensate for distortions due to frequency-dependent loss in transmission. One potential issue with the AWG is its temporal discretization of the sawtooth waveform, though numerically we will find that the sample clock of the utilized AWG is sufficiently fast that its effect is minimal. 

From the output of the AWG, the sawtooth wave is then directed into a broadband amplifier to amplify the signal, transmitted through cables, and then applied to the modulator. When designing this signal path there are two primary concerns. First the amplifier must have enough gain and saturation power so it can fully drive a $2\pi$ phase shift on the modulator. Having a modulator with a low $V_\pi$ is crucial to easily meet this requirement since achieving a high output power and simultaneously high bandwidth RF amplifier is difficult. Second, there will be frequency dependent losses through all components including the amplifier, cables, and the modulator itself which will introduce distortions into the signal. It is important to minimize these losses through through utilizing short, high efficiency cables and engineering a high-bandwidth modulator. When these are well engineered to be on the order of a few dB, the distortions from the modulator and cable can be effectively compensated via pre-distortion from the AWG. The amplifier, however, has a much sharper roll-off at its cutoff frequency that cannot be effectively compensated for, and thus it is what ultimately limits the achievable shift efficiencies particularly at larger frequency shifts (higher $f_{ST}$).

\subsubsection{Serrodyne Modeling}
To quantitatively understand the limitations of serrodyning, we model the effect of distortions of the sawtooth wave $\phi (t)$ on the frequency shift efficiency. We consider an electrical sawtooth wave with frequency $f_{ST}$ aiming to impart an optical frequency shift of $\Delta \omega = f_{ST}$ to the input CW wave. The primary figure merit we care about is the shift efficiency ($\epsilon$) into the target frequency bin as a function of $f_{ST}$. As $\phi (t)$ is periodic, the transmitted wave $E_{T}(t) = Ae^{j(\omega t + \phi (t))}$ is also periodic with period $1/f_{ST}$. Thus the frequency spectrum of the transmitted field is discrete , $$\|\mathcal{F}(E_{T}(t))\|^2 = \sum^{inf}_{n = -inf} A_n \delta(\omega + 2\pi nf_{ST})$$ where $\mathcal{F}$ is the Fourier transform operation. As the goal is to frequency shift into the n=1 bin, the efficiency can be trivially defined as $$\epsilon = A_1 / \sum_n {A_n} $$ Thus through this set of equations, the shift efficiency can be easily computed from a modeled $\phi (t)$.

In terms of modelling $\phi (t)$ we used a three step process. First $\phi (t)$ was assumed to be an ideal sawtooth wave with amplitude $2\pi$ and frequency $f_{ST}$ (10 GHz in this initial example), and from this $\|\mathcal{F}(E_{T}(t))\|^2$ and the efficiency $\epsilon$ of the frequency shift are computed (Fig \ref{TDSerrodyne}a). This is the "ideal" case and a shift of unity efficiency is expected. After this, we incorporate the effect of the digital emulation of the electrical waveform, namely temporal discretization. We temporally discretize the sawtooth wave at a sampling rate of $f_{AWG} \approx 100 GSa/s$ With this we can recompute $\|\mathcal{F}(E_{T}(t))\|^2$ and the efficiency $\epsilon$, and observe that although there is some penalty to efficiency due to this discretization, the sampling rate is high compared to $f_{ST}$ so this penalty is not particularly significant (Fig \ref{TDSerrodyne}b). 

\begin{figure}[ht]
\centering
{\includegraphics[scale=0.35]{SI 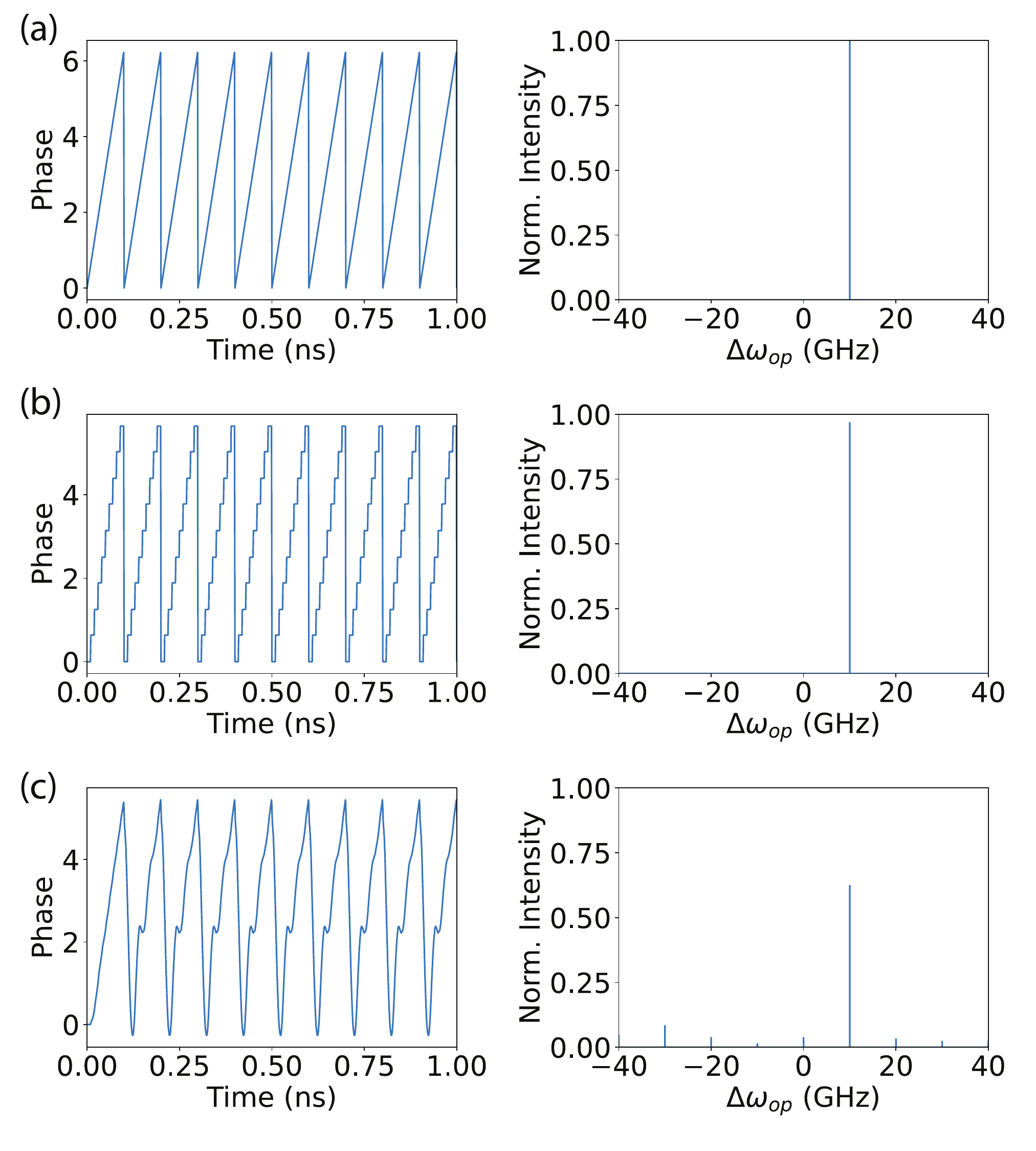}}
\caption{\textbf{Serrodyne Modeling.} Applied phase vs time (left) and corresponding optical frequency spectrum (right) for three different serrodyne models: (a) ideal, (b) temporally discrete, and (c) low pass filtered. The introduction of a finite analog bandwidth and a corresponding filter function induces a significant penalty in the achieved shift efficiency  }
\label{TDSerrodyne}
\end{figure}

Finally we want to model the effect of our system's finite analog bandwidth on $\phi (t)$ and the serrodyne efficiency. To do so we apply a filter function $H_{sys, eff} (f)$ to $\phi (t)$ where $H_{sys, eff} (f)$ is the effective transfer filter function of the system. Determining an effective $H_{sys, eff} (f)$ is somewhat challenging, as although the overall system function $H_{sys}$ is fairly simple to measure with standard laboratory equipment, the AWG's ability to perform pre-emphasis to partially eliminate distortions from this transfer function makes it more challenging to provide an effective $H_{sys, eff} (f)$.  As discussed previously, the actual transfer function of the system $H_{sys}$  includes a gradual decrease in transmission efficiency versus frequency due to a combination of variations in the passband of the amplifier as well as increased cable attenuation at higher frequencies. In addition to this, $H_{sys}$  has a much steeper cutoff at ~34 GHz due to cutoff of the amplifier's bandwidth (Fig \ref{AmpFit}). In order to provide a simple model for $H_{sys, eff}$ , we assume that pre-emphasis on the AWG is able to effectively pre-compensate the small variations below 30 GHz and only fails to compensate for the steeper rolloff form the amplifier beyond 34 GHz. Thus we approximate $H_{sys, eff} (f)$ as be a type II Chebyshev filter whose frequency cutoff and order are set such that the rolloff matches the measured system rolloff accurately (Fig \ref{AmpFit}). A type II Chebyshev filter is used due to the combination of steepness in response at the cutoff frequency, to accurately mimic the measured steep cutoff, and flatness in the passband to mimic ideal pre-compensation in the passband. We observe that the application of $H_{sys, eff} (f)$ to $\phi (t)$ leads to a significant distortion leading to a significant degradation in $\epsilon$, thus indicating this is the most significant practical penalty in the system (Fig \ref{TDSerrodyne}c). 

\begin{figure}[ht]
\centering
{\includegraphics[scale=0.4]{SI Figs/AmplifierFit.pdf}}
\caption{\textbf{Amplifier Fitting.} Measured transmission versus frequency of the system and corresponding filter used as a numerical approximation.}
\label{AmpFit}
\end{figure}

\subsubsection{Serrodyne Performance}

Using this model, we extract the shift efficiency $\epsilon$ as a function of optical shift frequency $\Delta \omega = f_{ST}$ (Fig \ref{SerrodyneEfficiency}). We observe we accurately match our measured serrodyne efficiencies (see main text Fig 4d). This both verifies the integrity of our model and indicates that in practice the analog bandwidth of the overall system, which is limited by the bandwidth of the broadband power amplifier, is the main limitation in the overall serrodyne efficiency. 

To understand how much improvement can be expected via serrodyning with state-of-the-art high-speed electronics, we use our validated model to estimate the efficiencies achievable with serrodyning with $f_{AWG} = 250$GSa/s and an analog bandwidth of 100 GHz \cite{Chen2017}. We observe even more significant improvements in the realized efficiencies are achievable, with shifts of efficiency greater than 50$\%$ possible up to 30 GHz and greater than 90 $\%$ up to 10 GHz . This demonstrates that electro-optic frequency serrodyning is a promising technique for overcoming the inhomogenous distribution of solid-state emitters. 

\begin{figure}[ht]
\centering
{\includegraphics[scale=0.4]{SI Figs/SerrodyneRolloff.pdf}}
\caption{\textbf{Serrodyne Efficiency Calculation.} Computed achievable serrodyne frequency shift efficiency as a function of shift amount for 35GHz and 100GHz analog bandwidths.}
\label{SerrodyneEfficiency}
\end{figure}

\subsection{Power Dissipation of Electro-Optic Shifting}
\label{PowerDissipationEOShifting}
One technical point of concern with electro-optic shifting is the amount of power dissipated on chip with the different methods for shifting. Too much dissipation can limit deployment of these techniques for a large number of memories or in a cryogenic environments.

For serrodyning, the power consumption can be fairly easily calculated as the modulator is terminated in a 50 $\ohm$ load. Thus the average power consumption is given by $P_{serr} = \frac{1}{T_{ST}}\int_0^{T_{ST}}v(t)^2/50 \Omega$ where $T_{ST} = 1/f_{ST}$ is the period of the sawtooth wave and the voltage across the modulator is given by $v(t) = V_\pi \cdot \phi(t) / \pi$. Evaluating this integral for $\phi(t) = 2\pi (t / T_{ST} - 1/2 )$ (one period of a sawtooth wave) we arrive at : $$P_{serr} = \frac{1}{3}\cdot\frac{V_\pi^2}{50\Omega}$$. As this is total power consumption, power dissipation is given by the proportion of this power which is lost when transmitted across the modulator's electrodes. While doing an exact calculation for this will be electrode and frequency dependent, due to the relatively high loss of the modulator's electrodes particularly at high frequencies, a large proportion of the incident power will likely be absorbed. Thus for purposes of an upper bound, we can assume the on-chip dissipated power is equivalent to $P_{serr}$. For the modulator demonstrated in this work, $V_{\pi} \approx 1.75V$ giving a total power consumption of $\sim$20 mW. We note this is likely a pessimistic estimate, particularly at smaller $f_{ST}$.

For shearing, although peak to peak voltages much greater than $V_\pi$ are required, the narrow-band and relatively low frequency nature of the electrical waveform ensures that the modulator electrode can be used in a lumped element configuration, significantly decreasing the power dissipation. To calculate this dissipation, We model the modulator as a lumped element R-C circuit with resistance $R_{mod}$ and capacitance $C_{mod}$. In the limit where the frequency of the applied bias $f_{shear} << R_{mod}C_{mod}/(2\pi)$ the power dissipated across the resistance $R_{mod}$ is given by $$P_{shear} = R_{mod}\cdot(2\pi\cdot f_{shear} C_{mod} V_{RMS})^2$$. For the shearing demonstration in this work, $R_{mod} \approx 5\Omega$, $f_{shear} = 250$ MHz, $C_{mod} = 2$pF (based on modeling of the measured electrode parameters), $V_{RMS} = \sim 35$V so we obtain $P_{shear} \approx 60$ mW.

 \subsection{Segmented Electrode Parameters}

 In order to perform electro-optic frequency conversion, it is important that the electro-optic response of the modulator is maximized at larger frequencies, particularly for shearing. In our previous work \cite{Renaud2023}, our modulator utilizing a standard co-planar waveguide design had a significant initial rolloff in EO response due to a large impedance mismatch, as the center trace had to be made small in order to match the RF index to the group velocity of the waveguide ($\approx$ 2.4). To overcome this, we chose to utilize segmented electrodes for our modulator. These segmented electrodes increase the microwave index, allowing us to better match the microwave index with the waveguide index (with the addition of a thicker buried oxide of 7 $\mu$m). In addition, this new design has the additional benefit of significantly decreased RF loss due to the breaking of the inner conductor current path (see Table \ref{tableParams}). These factors combine to enable an enhanced EO response.

 \begin{table}[ht]
 \centering
     \begin{tabular}{||c c c||}
         \hline
         & \textbf{Regular} & \textbf{Segmented} \\
         \hline
         \textbf{Impedance ($\Omega$)} & 36 & 40 \\
         \hline
         \textbf{RF Loss (dBcm$^{-1}$GHz$^{-1/2}$)} & 1.31 & 0.73\\
         \hline
         \textbf{RF Index}  & 2.41 & 2.4\\
         \hline
     \end{tabular}
     \caption{\textbf{Modulator electrode parameters} for our previous work \cite{Renaud2023} and this work}
    \label{tableParams}

 \end{table}

 \begin{figure}[ht]
\centering
{\includegraphics[scale=0.45]{SI Figs/EO_vs_freququency_segmented.pdf}}
\caption{\textbf{Simulated electro-optic response} as a function of frequency for regular and segmented electrode modulator, demonstrating a significantly improved rolloff for the segmented design.} 
\label{SegmentedVsReg}
\end{figure}

\section{System Considerations}
\subsection{Photonic Integrated Circuit Schematic and Explanation of Requirements}
\label{LNOIArchitecture}
A technical diagram of our proposed TFLN architecture is shown in Figure \ref{TDSerrodyne}. An input CW laser light is first transmitted through a laser control unit (LCU) that is used to generate and shape the temporal and spectral profile of the optical pulse that  control the quantum memory (QM).   LCU consists of  an amplitude modulator (AM) that carves  the optical pulse out of CW light, and a phase modulator (PM) that enables frequency shifting of a common seed laser light to address different, inhomogeneously broadened, QMs. Next,  optical pulse is transmitted to  the QM , and spin-photon entanglement is generated through one of many  schemes \cite{Nguyen2019, Chan2023, Hensen2015}. The  spin-entangled photon is then passed  through a Multiplexing Unit (MU) that consists of a PM, switch network, fiber couplers and on-chip photodetectors. PM is used  to shift its frequency of spin-entangled photon to a common frequency shared by all of the individual quantum memories within a given memory node, thereby overcoming the inhomogeneous distribution challenge \cite{Evans2016, Machielse2019}. The common-frequency photons are routed using a switch network either to an on-chip photon detector \cite{Colangelo2024} to enable high-efficiency on-chip heralded entanglement between memories within the same node, or an output coupler so that photons from different memories can be efficiently multiplexed across the wider fiber network.  

\begin{figure}[ht]
\centering
{\includegraphics[scale=0.7]{SI Figs/SI_schematic.pdf}}
\caption{\textbf{Schematic of LNOI multiplexed quantum node.} }
\label{TDSerrodyne}
\end{figure}

\subsection{Power Consumption and Dissipation of Switching}
One major component of the power consumption and dissipation of the system is that of the switching elements. As the switches are capacitively loaded (not 50 $\Omega$ terminated) they can be modeled as an $RC$ network where $R_{tot}$ is given by the total resistance of network, given by the sum of the source and system resistance $R_{sys}$ and the device resistance $R_{sw}$, and the capacitance is given by the device capacitance. The dynamic power consumption is given by $P_{cons} = fCV^2$: assuming a switching frequency $f$ and with a voltage $V = V_\pi$ of the switch. The actual dynamic power dissipated on chip is given by $P_{diss} = R_{sw} / R_{tot}\cdot P_{cons}$, as it is a simple resistive power divider between the system and switch resistances.

For the switch presented in the main text, we estimate $C = 1$pF, given by estimation of the distributed capacitance using the measured transmission line parameters and telegrapher's equations, and $R_{sw} = 5 \Omega$. Assuming $R_{sys} = 50 \Omega$ and given the measured $V_\pi = 1.5$V we estimate a dynamic power consumption of $\approx 200 \mu W$ and an on chip power dissipation of $\approx 20 \mu W$ for a switching frequency of $f = 100$ MHz. 

This can likely be substantially improved through only minor modifications to the geometry. First, the geometry can be shrunk to 1mm. Although this will increase $V_\pi' = 7.5V$, we estimate a decreased $R_{sw}' = 1 \Omega$ and $C = 0.2$pF. Moreover, we can artificially increase the system resistance $R_{sys}$ so that a larger proportion of power consumption is dissipiated at the system as opposed to on-chip. This is important for thermal management of the chip, particularly in cryogenic environments. The cost of increasing $R_{sys}$ is the maximum frequency response of the system decreases as it is set by the RC time constant $R_tot \cdot C$, but since only a moderate switching frequency of 100 MHz or so is desired, $R_{tot}$ can be increased significantly without a performance penalty. Thus we assume $R_{sys} = 5k\Omega$. This optimized set of parameters enables an on-chip power power dissipation of $\sim 0.2 \mu W$. We note that the estimated total power consumption (as opposed to on-chip dissipation) of this optimized geometry actually increased to $\sim 1$ mW  primarily due to the larger $V_\pi'$ due to the shorter length. This thus indicates a partial trade-off between in the switch performance between total power consumption, on-chip power dissipation, and footprint that has to be evaluated on a system by system basis. 

We also note that the on-chip power dissipation, and likely the total power consumption of the switches are considerably smaller than the power consumption needed for electro-optic frequency shifting (see Section \ref{PowerDissipationEOShifting}) and thus that will ultimately dominate the on-chip power dissipation for any full-scale realization of our system. 

\subsection{Co-packaging PIC with Quantum Memories}

Heterogeneous integration of a quantum memory module with an active PIC (in this case TFLN) has often been considered as a solution to integrate active switching functionality into a passive quantum memory platform such as diamond. Although there have been a variety of works showing capabilities to achieve this and the advantage of this approach is clear, namely minimizing the net insertion loss at the quantum memory - active photonics interface, this direct integration approach has numerous practical problems. These practical issues generally stem from the fact that the quantum memories have stringent cryogenic requirements, with state-of-the art quantum memories needed to operate at 1K or even below to achieve good coherence properties. This cryogenic requirement is largely not compatible with the thermal and size requirements of the PIC. More specifically: 

\begin{enumerate}
\item Power dissipation of switching is not insignificant. We estimate our current switches dissipate 20 $\mu$W of power. For $\sim$1000s of memories and thus switches required to maximize a quantum link's entanglement rate (see \ref{MultiplexingTheory}), that corresponds to 10s mW of dissipated power, which is a similar order of magnitude to the cooling power of 4K systems. In addition, the implementation of electro-optic frequency shifting will lead to the dissipation of 10s of $mW$ of power per memory, indicating that cooling power on the order of $W$ is required. In addition, these estimates do not include resistive heating via the cabling which likely will be significant.

\item Electrical control of the switches requires extensive electrical wiring. Assuming complete independent control of each modulator is required to achieve the desired multiplexing and local operation scheme, each modulator requires an independent control line. As a full PIC will likely need to include 1000s of modulators in order to achieve optimal temporal multiplexing,  this requires 1000s of electrical wires. The heat conductivity of wires is large, and thus adding wiring from the PIC at the base of the cryostat to a controller in the external ambient environment could easily lead to a prohibitive heat load. Potentially integration of the CMOS controller within the cryostat either at the base plate or a higher temperature stage could solve this, though cryogenic compatible CMOS technology is still at early stages.

\item The footprint of the modulators is comparatively large, with lengths on the order of -mm to even -cm, and lateral density of 100s of $\mu$m limited by optical and electrical cross-talk. Thus the total space in the cryostat occupied by these switches is quite large.

In contrast, 2D fiber arrays with pitches down to 82 $\mu$m are commercially available. Thus a more practical alternative could be to utilize high-density optical packaging and leave the active PIC outside of the cryostat. In particular, in architectures which coupling loss can be rolled into total channel loss between quantum nodes, dB level excess losses that might be incurred from this massive coupling approach can be tolerated if it enables enhanced scalability. Alternatively a hybrid  solution, with limited switching in the cryostat both in number of stages and connectivity and the rest in an external PIC at ambient temperature, might provide a best of both worlds solution.

\end{enumerate}

\subsection{Multiplexing Theory}
\label{MultiplexingTheory}
We construct a simple model of remote entanglement generation in order to understand the benefits of multiplexing enabled by the TFLN PIC. 

Assume there are two nodes separated by a distance $L$ each with $N$ quantum memories with a measurement station in the middle at a distance $L/2$ from either node. At the start of the protocol which is repeated at a clock rate $f_{clk}$, each node probabilistically emits a spin-entangled photon with duration $t_{photon}$. The photons then are sent to a middle measurement node in which a partial (1 photon) or full (2 photon) bell state measurement occurs on the photon(s). The successful heralded detection of 1 (2) photons indicates entanglement was successfully generated. The probability that a given attempt leads to a heralded entanglement event is given by $p_{entang}$

The overall entanglement rate $R_{entang}$ is given by the product $R_{entang} = p_{entang} \cdot f_{clk}$. In the case of $N=1$ memories per node and no multiplexing, the maximum possible clock rate $f_{clk}$ is determined by the time between a protocol attempt and determining whether that entanglement generation was successful. This is since there is only a single memory per node, and thus another entanglement generation attempt cannot be started as it would require re initializing that memory and thus eliminating entanglement before it is even generated. This temporal delay  is determined by the combination of time it takes for the photon to arrive from the two quantum memory nodes at the central measurement node, and the time it then takes for the classical signal to propagate back indicating whether entanglement was successfully generated  Thus we have $1/f_{clk, N=1} = 2\cdot \frac{L}{2} \cdot \frac{1}{c'} = \frac{L}{c'}$ where $c'$ is the effective speed of light in the propagation medium. 

When the number of memories is increased $N > 1$, the clock rate can be increased through temporal multiplexing of memories on the channel. After a photon is entangled with one memory and sent and the node is waiting for the result, an entanglement attempt can be performed with the second memory, and so on until all of the memories at a given node have been utilized and are waiting for their respective results. Thus although the delay time between entanglement attempt and determining whether it was successful is still the same ($\frac{L}{c'}$), the effective clock rate is increased by the number of memories $n$: $1/f_{clk, N=n} = \frac{L}{c'\cdot n}$. The limit of temporal multiplexing is ultimately set by the temporal-length of the spin-entangled photon, as only one photon can exist within the channel at a given time (assuming a matched polarization and frequency). This thus sets the maximum clock rate $1/f_{clk, max} = t_{photon}$. 

 $p_{entang}$ can be broken down into the product of four individual probabilities: $p_{photon}$: the probability a spin-entangled photon is generated and coupled into single mode fiber ($p_{photon} = p_{photon, gen} \cdot p_{photon\rightarrow fiber}$), $p_{PIC->fiber}$ the probability the photon is transmitted through the PIC into the output fiber, $p_{fiber}$ which is the probability the photon is successfully transmitted across the fiber channel (including the efficiency of frequency conversion from visible to telecom wavelengths - $p_{fiber} = p_{vis\rightarrow telecom} \cdot p_{fiber, T}$), and $p_{protocol}$ which encompasses the probability of entanglement success due to the probabilistic nature of heralded entanglement generation itself. We then see the overall entanglement success probability is $p_{entang} = (p_{photon} \cdot p_{PIC->fiber} \cdot p_{fiber})^{n_{photon}} \cdot p_{protocol}$ where $n_{photon}$ is the number of photons required to be detected to herald entanglement which is protocol dependent.

$p_{PIC->fiber}$ depends on the exact architecture and independent device losses of the PIC and most notably will increase with larger number of memories $N$ as a larger switch network is required to multiplex a larger number of quantum memories. Using the architecture we propose in this work (and described in section \ref{LNOIArchitecture}) we determine that the full insertion loss of the PIC is given by $IL_{PIC} = 2\cdot IL_{coupling} + (n_{switch})\cdot IL_{switch} + IL_{pm} + IL_{\Delta \omega_{Op}}$. We note that we assume that the PIC and the quantum memory bank are not co-packaged and thus we take double the on-off chip insertion loss penalty. The number of switches $n_{switch}$ is dependent on the number of memories that are multiplexed: $n_{switch} = log_2N + 1$.

To highlight the benefits of multiplexing we compute $R_{entang}$ as a function of $N$. In Table \ref{MultiplexParams} we list the parameters used for this calculation. The insertion losses for the computation of $p_{PIC->Fiber}$ were all extracted from the measurements presented in this work. The rest of the parameters assumed use of the silicon-vacancy center in diamond where applicable.

For the comparison case of a non multiplxed quantum node, we fix $N = 1$,  $p_{PIC} = 1$ and repeat the calculation. We observe that in spite of the additional insertion losses required to implement multiplexing, the losses of the devices in our platform are sufficiently low such that the benefits of multiplexing outweigh the excess losses. We thus predict over a 100x gain in entanglement rate through multiplexing with our LNOI PIC.

 \begin{table}[ht]
     \begin{tabular}{||p{0.15\linewidth} | p{0.3\linewidth} | p{0.2\linewidth} |p{0.25\linewidth}||}
         \hline
         \textbf{Parameter} & \textbf{Description} & \textbf{Value} & \textbf{Notes}\\
         \hline
         L & Node-Node Distance & 20 km & \\
         \hline
         c' & Group velocity of fiber  & $2\cdot 10^8$ m/s & \\
         \hline
         t\textsubscript{photon}  & Photon Lifetime & 50 ns & \cite{Nguyen2019}.\\
         \hline
         n\textsubscript{photon} & Number of Photon Event & 1 &\\
         \hline
         p\textsubscript{photon} & Probability of Spin-Photon Generation & 0.493 $\times$ 0.4 & Product of demonstrated spin-photon gate efficiencies (Ref \cite{Bhaskar2020}) times high $\mu$ weak coherent source. \\
         \hline
         p\textsubscript{protocol} & Protocol success probability & 0.25 & Protocol probability for 1-photon heralded protocol \cite{Humphreys2018} \\
         \hline
         p\textsubscript{photon->fiber} & Efficiency of single photon coupling into single mode fiber & 0.9 (-0.4dB) & \cite{Zhang2023} \\
         \hline
         p\textsubscript{vis$\rightarrow$telecom} & Efficiency of conversion from visible to telecom wavelengths for long-distance fiber transmission & 0.5 (-3dB) & \cite{Leent2020} \\
         \hline
         p\textsubscript{fiber, T} & Transmission efficiency through length $L/2$ fiber & 0.5 (-3dB) & 0.3 dB/km \\
         \hline
         IL\textsubscript{coupling} & Insertion loss of coupling into LNOI chip & 0.8 (-1dB) & This work \\
         \hline
         IL\textsubscript{switch} & Insertion loss of LNOI switch & 0.9 (-0.4dB) & This work \\
         \hline
         IL\textsubscript{pm} & Insertion loss of LNOI phase modulator & 0.85 (-0.7dB) & This work \\
         \hline
         IL\textsubscript{$\Delta \omega\textsubscript{Op}$} & Frequency shifting efficiency & 0.9 (-0.5dB) & This work \\
         \hline
     \end{tabular}
     \caption{\textbf{Parameters used in multiplexing system calculation.} }
     \label{MultiplexParams}
 \end{table}

\newpage

\bibliography{sn-bibliography-upd}